\documentclass[preprintnumbers,amsmath,amssymb,twocolumn,pra]{revtex4-1}
\usepackage{graphicx,comment}% Include figure files
\usepackage{epstopdf}
\usepackage{dcolumn}% Align table columns on decimal point
\usepackage{bm}% bold math
\usepackage[T1]{fontenc}
\usepackage[latin1]{inputenc}
\usepackage{enumerate}
\usepackage{fancyhdr,color}
\usepackage{verbatim}

\def\Pr{\mathrm{Pr}}

\def\one{{\mathchoice {\rm 1\mskip-4mu l} {\rm 1\mskip-4mu l} {\rm
1\mskip-4.5mu l} {\rm 1\mskip-5mu l}}}
\newcommand{\sket}[1]{|#1\rangle}
\newcommand{\sbra}[1]{\langle#1|}
\newcommand{\ket}[1]{\left| #1\right\rangle}        % ket vector
\newcommand{\bra}[1]{\left\langle #1\right|}        % bra vector
\begin{document}
%\preprint{APS/123-QED}

\title{Improved Bounds for Eigenpath Traversal}

\author{Hao-Tien Chiang}
\email{lewis.prometheus@gmail.com}

\affiliation{
University of New Mexico \\
Albuquerque, New Mexico 87185, USA \\
}
\author{Guanglei Xu}
\email{glxu.leo@gmail.com}

\affiliation{
University of Pittsburgh \\
Pittsburgh, PA 15260, USA \\
}

\author{Rolando D. Somma}
\email{somma@lanl.gov}

\affiliation{
Los Alamos National Laboratory \\
Los Alamos, New Mexico 87545, USA
}

\begin{abstract}
We present a bound on the length of the path defined by the ground states
of a continuous family of Hamiltonians in terms of the  spectral gap $\Delta$. 
We  use this bound to obtain a significant improvement over the cost of recently proposed methods for 
quantum adiabatic state transformations and eigenpath traversal. 
In particular, we prove that a method based on evolution randomization, which is a simple extension of adiabatic quantum computation,
has an average cost of order $1/\Delta^2$, and a method based on fixed-point
search, has a maximum cost of order $1/\Delta^{3/2}$. Additionally, if the Hamiltonians satisfy a 
frustration-free property, such costs can be further improved to order $1/\Delta^{3/2}$ and $1/\Delta$,
respectively. Our methods offer an important advantage over adiabatic quantum computation
when the gap is small, where the cost is of order $1/\Delta^3$. 
\end{abstract}

\date{\today}% It is always \today, today,

\maketitle

%%%%%%%%%%%%%%%%%%%%%%%%%%%%%%%%%%
%%%%%%%%%%%%%%%%%%%%%%%%%%%%%%%%%%
\section{Introduction}
\label{sec:intro}
%{\color{red} can we be more precise?}
Numerous problems in quantum information,  physics and optimization, can be solved by preparing
the low energy or other eigenstate of a Hamiltonian~(cf. \cite{apolloni:qa89,finnila:qc1994a,kadowaki:qc1998a,farhi:qc2001a,sachdev_2001,santoro:qa06,somma_quantum_2008,perez_PEPS_2008,das:qa08,schwarz_peps_2011}).  
On a quantum system (e.g., an analogue quantum computer), such an eigenstate
can be prepared by smoothly changing the interaction parameters of the controlled Hamiltonians
under which the system evolves. That is the  idea of adiabatic quantum computation (AQC),
which relies on the adiabatic theorem~\cite{Born-Fock:adiabatic,messiah_1999} to assert that, at any time,
the evolved state is sufficiently close to an eigenstate of the system that is continuously related to the final one.

The importance of AQC for quantum speedups was  demonstrated in several examples~(c.f., \cite{farhi_quantum_2002,roland_quantum_2002,somma_quantum_2008,Hen_period_2013}). 
In particular, AQC is equivalent to the standard circuit model of quantum computing, implying that
 some quantum speedups obtained in one model may be carried to the other
using  methods that map quantum circuits to Hamiltonians and vice versa~\cite{aharonov_adiabatic_2007,mizel_equivalence_2007,oliveira_adiabatic_2008,aharonov_line_2009,somma_Feynman_2013,
cleve_query_2009,wiebe_product_2010,childs_efficient_2010,BCS2013}.
In AQC, we assume access to Hamiltonians $H(s)$, $0 \le s \le 1$,
that have  non-degenerate and continuously related eigenstates $\ket{\psi(s)}$.
The goal is to prepare $\ket{\psi(1)}$ from $\ket{\psi(0)}$, up to some small approximation
error $\varepsilon$, by increasing $s$ from 0 to 1 with a suitable time schedule.
The cost of the algorithm in AQC is determined by the total evolution time, $T$.
This time depends on properties of the Hamiltonians used in the evolution, 
such as their rate of change or spectral gaps.
In particular, a commonly used and rigorous quantum adiabatic approximation 
provides an upper bound to the cost given by (cf. \cite{jansen_bounds_2007,jordan_thesis_2008})
\begin{align}
\label{eq:AA}
T_{\rm AQC} = \kappa \max_s \left [ \frac{ \| \ddot H \|} {\varepsilon \Delta^2}, \frac{ \| \dot H \|^2} {\varepsilon \Delta^3} \right] \; .
\end{align}
That is, increasing $s$ according to, for example, $s(t)=t/T_{\rm AQC}$,
suffices to prepare the final eigenstate from the initial one within    error $\varepsilon$.
(The cost will be $T=T_{\rm AQC}$ for such a schedule.) $\kappa$ is a constant and
 $\Delta$ is the spectral gap of $H$, that is,
the smallest (absolute) difference between the eigenvalue of $\ket{\psi(s)}$ and any other eigenvalue.
%$s(t)=t/T_{\rm AQC}$ will be a slowly-varying  parameter when $\Delta$ is small. 
Unless  stated otherwise,  all quantities, states, and operators depend on $s$, and all derivatives are with respect to $s$,
e.g., $\dot X = \partial X / \partial s$ and
$\ddot X = \partial^2 X/\partial s^2$. For an operator or matrix $X$ and state $\ket \phi$ on a $d$-dimensional complex Hilbert space, 
 $\| X \|$ denotes the spectral norm and $\| \ket \phi \|$ denotes the Euclidean norm.
 We remark that the bound in Eq.~\eqref{eq:AA} is actually tight in the sense that there 
 exist examples (e.g., Rabi oscillations, c.f.~\cite{Marzlin_Inconsistency_2004,Amin_Consistency_2009}) for which the total cost of the adiabatic evolution is also lower bounded
 by a quantity of order $\| \dot H \|^2/\Delta^3$, and $\| \dot H \|^2/\Delta^3 > \| \ddot H \|/\Delta^2$
 in such examples.

 %Up to an arbitrarily small error, the quantum evolution prepares the final state $\ket{\psi(1)}$
%from the initial state $\ket{\psi(0)}$, i.e., it transforms the eigenstate
%of $H(0)$ into the eigenstate of $H(1)$.

A drawback with rigorous quantum adiabatic approximations
is that the dependence of $T_{\rm AQC}$ on the gap
is rather poor, specially when $\Delta \ll 1$.   
Also, the bound given by Eq.~\eqref{eq:AA} could  imply a large overestimate of the actual  cost
needed to prepare the final eigenstate in some cases. 
For these reasons, other methods for traversing the 
eigenstate path, which differ from AQC
but have a better cost dependence on the gap, were recently proposed~\cite{wocjan_speed-up_2008,boixo:qc2009a,boixo:qc2010a}.
One such method~\cite{boixo:qc2009a} is based on evolution randomization to implement
a version of the quantum Zeno effect and simulate projective measurements
of $\ket{\psi(s)}$. The main and only difference between this ``randomization method'' (RM)
and AQC is that, rather than choosing the schedule $s(t)=t/T_{\rm AQC}$ for the evolution,
$s(t)$ is randomly chosen according to a probability distribution that depends on the gap and the approximation error.
Another method~\cite{boixo:qc2010a} also traverses the eigenstate path by making projective measurements
of $\ket{\psi(s)}$, but each measurement is implemented using the so-called
phase estimation algorithm~\cite{kitaev_quantum_1995} and Grover's fixed-point search technique~\cite{grover_different_2005}. The method in Ref.~\cite{boixo:qc2010a} requires knowing the eigenvalue of $\ket{\psi(s)}$,
but this can be {\em learned} as the path is traversed.

The (average) cost $T$, or total  time of evolution under the $H(s)$, of the previous methods for eigenpath traversal 
depends not only on the spectral gap but also on the 
eigenstate path length, $L$. This is simply the length defined
in complex Hilbert space: $L = \int_0^1 ds \| |\dot \psi \rangle \|$.
For   error $\varepsilon<1$, the cost is upper bounded by
\begin{align}
\label{eq:costEPT1}
T_{\rm EPT} =\kappa' \frac {L^c \log (L/\varepsilon)} {\varepsilon \min_s \Delta}  \; ,
\end{align}
with $c=1,2$ depending on the method and $\kappa'$ a constant. Having an explicit
dependence in the path length is important for those cases
in which $L$ can be  bounded independently of the gap. 
This observation was used in Ref.~\cite{somma_quantum_2008}
to prove a quantum speedup of the well-known
simulated annealing method used for optimization~\cite{kirkpatrick_SA_83} (Sec.~\ref{sec:QSA}).
For many hard optimization problems, $\Delta$ decreases 
exponentially in the problem size while $L$
increases only polynomially. Then, $T_{\rm AQC} \gg T_{\rm EPT}$ for these cases
and the methods in Refs.~\cite{wocjan_speed-up_2008,boixo:qc2009a,boixo:qc2010a}
may be used to prepare the final eigenstate with lower cost than the adiabatic method.

We remark that the  upper bound   of Eq.~\eqref{eq:costEPT1}
can only be achieved for a uniform parametrization, under which the eigenstate satisfies $\| \sket{\dot \psi } \| =L$,
independently of $s$. This is a strong requirement that will not be satisfied
in general.
% merely computing $L$ from $\| \dot H \|$, $\| \ddot H \|$ and $\Delta$ could be impossible. 
We then considered an upper bound $L^* \ge L$, which can be easily computed from known properties of the Hamiltonians,
and used such a bound to obtain the corresponding  $T_{\rm EPT}$ 
in Refs.~\cite{boixo:qc2009a,boixo:qc2010a} (i.e., by replacing $L \rightarrow L^*$).
When   $\| \dot H \|$ and $\Delta$ are known, a commonly used
path length bound is
\begin{align}
\label{eq:Lbound0}
L^* = \max_s \frac{\| \dot H \|}{\Delta}  \; .
\end{align}
Such a bound follows easily  from the eigenvalue equation,
which can be used to obtain $\| |\dot \psi \rangle \| \le \| \dot H \| / \Delta$
~\cite{NoteEPT-bounds}. Equations~\eqref{eq:costEPT1} and~\eqref{eq:Lbound0}  give an upper bound for the cost of the eigenpath traversal method as
\begin{align}
\label{eq:costEPT2}
T_{\rm EPT} = \kappa' \max_s \frac{\| \dot H \|^c}{\varepsilon \Delta^{c+1}} \log (\| \dot H \|/(\varepsilon \Delta)) \; .
\end{align}
$c=2$ for the RM and $T_{\rm EPT}$ can be larger than $T_{\rm AQC}$ when the parametrization
is different from the uniform one.
Thus, the advantage of the RM over the adiabatic method is unclear in this case from the above upper bounds:
 both, $T_{\rm AQC}$ and $T_{\rm EPT}$, depend on $1/\Delta^3$.

A main goal of this paper is to obtain better bounds for the cost of the methods of Refs.~\cite{wocjan_speed-up_2008,boixo:qc2009a,boixo:qc2010a} in terms of the spectral gap, the error,  $\| \dot H \|$ and $\| \ddot H\|$,
giving special emphasis to the RM described in Ref.~\cite{boixo:qc2009a}.
Such quantities, or bounds of, are assumed to be known.
The reason why we focus more on the RM than other methods for eigenpath
traversal is due to its simple connection with AQC. The other methods not only require evolving with the Hamiltonian,
but also require implementing other operations such as those for the quantum Fourier transform in the phase
estimation algorithm. Nevertheless, some of our results can also be used to improve the cost of those other methods as well.

Our manuscript is organized as follows.
 In Sec.~\ref{sec:pathlength},
we  present an improved bound on the path length
where, ignoring other quantities, $L^*$ is of order $1/\sqrt \Delta$ if $\ket{\psi}$ is the ground state of $H$.
We study this bound for general Hamiltonian paths and focus also on those Hamiltonians that are {\em frustration free},
due to their importance in condensed matter theory~\cite{perez_PEPS_2008,feiguin_renorm_2013}, optimization~\cite{somma_thermod_2007}, and quantum information~\cite{bravyi_stoquastic_2009,beaudrap_frustfree_2010,somma_gap_2013}.
Then, in Sec.~\ref{sec:RMbound}, we use the improved bound to obtain
an average cost for the RM of order $1/\Delta^2$, which is much smaller than
$T_{\rm AQC}$ when $\Delta \ll 1$. In Sec.~\ref{sec:weakmeasurement}
we improve the analysis of Ref.~\cite{boixo:qc2009a} about the cost scaling with the error
and show that the logarithmic factor present in Eqs.~\eqref{eq:costEPT1} and \eqref{eq:costEPT2} for the RM is unnecessary.
In Sec.~\ref{sec:applications} we apply our  results to two important problems in quantum computation,
namely the preparation of projected entangled pair states~\cite{verstraete_peps_2006} (i.e., generalized matrix product states
or PEPS) and
the quantum simulation of classical annealing processes~\cite{kirkpatrick_SA_83,somma_quantum_2008}.
We use the results for frustration-free Hamiltonians and show that the RM
has an average cost of order $1/\Delta^{3/2}$ for the preparation of PEPS, while
the method based on fixed-point search 
has cost of order $1/\Delta$ (up to a logarithmic correction).
 We conclude in Sec.~\ref{sec:conclusions}

%%%%%%%%%%%%%%%%%%%%%%%%%%%%%%%%%%%%%%%%%%%%%%%%%
\section{The path length}
\label{sec:pathlength}
The path length of a continuous and differentiable family of unit states $\{ \ket{\psi(s)} \}$, $0 \le s \le 1$, is
\begin{align}
\nonumber
L=\int_0^1 ds \| \sket{\dot \psi} \| \; .
\end{align}
The global phase of $\ket{\psi}$ is set so that $\langle \psi \sket{\dot \psi}=0$.
$\ket{\psi}$ is a non-degenerate eigenstate of $H$ and, without  loss of generality, we assume that
the eigenvalue is 0.
Then, $\sket{\dot \psi}=-H^{-1}\dot H \ket{\psi} $, where $H^{-1}$ has only support in the subspace orthogonal
to $\ket \psi$. An upper bound of $\max_s(\|\dot H\|/\Delta) $ on $L$ simply follows.
Such a bound is commonly used when deriving adiabatic approximations.

Remarkably, if the state path is two times differentiable and $\ket{\psi}$
is the ground state of $H$ (i.e., the eigenstate with lowest eigenvalue), a tighter bound on $L$ in terms of the gap can be obtained.
According to the Cauchy-Schwarz inequality,
\begin{align}
\label{eq:Lbound1}
L^2 \le \int_0^1 ds \; \| \sket{\dot \psi} \|^2 \; .
\end{align}
By differentiation of $H \ket \psi=0$, in Appendix~\ref{app:rateofchange}
we obtain
\begin{align}
\label{eq:Lbound2}
\| \sket{\dot \psi} \|^2 \le \frac { 1}{2 \Delta} \bra \psi \ddot H \sket{ \psi} \; ,
\end{align}
see Eq.~\eqref{eq:dotpsibound}.
Equations~\eqref{eq:Lbound1} and~\eqref{eq:Lbound2} yield
\begin{align}
%\label{eq:Lbound4}
\nonumber
L^2 \le   \int_0^1 ds \; \frac 1 {2 \Delta} \bra{\psi} \ddot H \ket {\psi} \; .
\end{align}
If the lowest eigenvalue is $E \ne 0$, then
\begin{align}
\label{eq:Lbound5}
L \le L^* =\left(  \int_0^1 ds \; \frac 1 {2 \Delta} \bra{\psi} \ddot H -\ddot E \ket {\psi}  \right)^{1/2}\; .
\end{align}
Equation~\eqref{eq:Lbound5} is our main result; its applications  
to eigenpath traversal will be discussed below.

%%%%%%%%%%%%%%%%%%%%%%%%%%%%%%%%%%
\subsection{General interpolations}
\label{sec:generalinterpolations}
In general, because $\bra{\psi} \ddot H -\ddot E \ket {\psi} \ge 0$, the {\em rhs} of
Eq.~\eqref{eq:Lbound5}  can be bounded so that
\begin{align}
\nonumber
L^* & \le \max_s \sqrt{ \frac{\| \ddot H \| -(  \dot E(1) -  \dot E(0) ) } {2\Delta}}  \\
\nonumber
& \le \max_s \sqrt{ \frac{\| \ddot H \| +2 \| \dot H \|} {2 \Delta}}  \; .
\end{align}
For eigenpath traversal, quantities such as $\| \dot H \|$ and $\| \ddot H \|$
are usually bounded by a polynomial on the problem size, while the spectral
gap $\Delta$ can be exponentially small for hard instances.

%%%%%%%%%%%%%%%%%%%%%%%%%%%%%%%%%%
\subsection{Linear interpolations}
\label{sec:linearinterpolation}
A commonly used Hamiltonian path is given by the linear interpolation
of two Hamiltonians, that is, $H=(1-s)H_0 + sH_f$. Here, $H_0$ and $H_f$
are the initial and final Hamiltonians, respectively.
In this case, 
\begin{align}
\nonumber
L^* & \le \max_s \sqrt{\frac{\dot E(1) - \dot E(0)}{2\Delta}} \\
\nonumber
& \le \max_s \sqrt{ \frac{\| \dot H \|}{ \Delta }} \; .
\end{align}

%%%%%%%%%%%%%%%%%%%%%%%%%%%%%%%%%%%%%%%%
\subsection{Frustration-free Hamiltonians}
\label{sec:frustrationfree}
A Hamiltonian $H=\sum_k \Pi_k$ is said to be frustration-free
if any ground state $\ket{\psi}$ of $H$ is also a ground state of every $\Pi_k$.
Typically, $\Pi_k$ corresponds to local operators and we can assume that
$H \ket{\psi}= \Pi_k \ket \psi=0$ for all $k$, and $\Pi_k \ge 0$.

For frustration-free Hamiltonians, the {\em local} bound on the 
rate of change of the state in Eq.~\eqref{eq:Lbound2} applies directly because $E=0$, and then
\begin{align}
\label{eq:pathlengthFF}
L^* \le \max_s \sqrt{ \frac{\| \ddot H\|} {2 \Delta}} \; .
\end{align}

%%%%%%%%%%%%%%%%%%%%%%%%%%%%%%%%%%%%%%%
\section{Improved bounds for the randomization method}
\label{sec:RMbound}
\begin{figure}[ht]
  \centering
  \includegraphics[width=8.5cm]{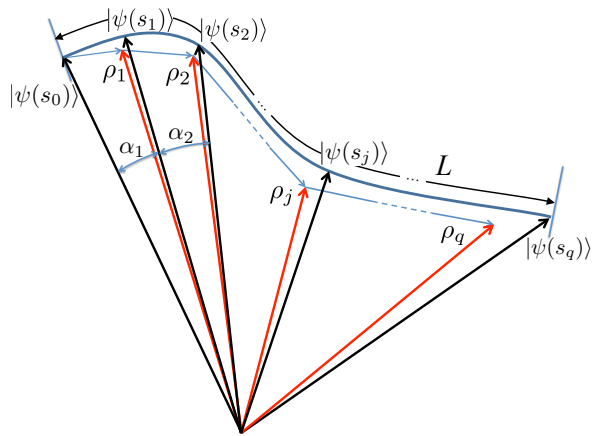}
  \caption{Basic steps of the RM and state representation. At the $j$ th step, the RM prepares
  the mixed state $\rho_j$ (represented by a red arrow) that has large probability of being in $\ket{\psi(s_j)}$ (
  represented by a black arrow)
  after measurement. The preparation of $\rho_j$ is done by evolving $\rho_{j-1}$ with the Hamiltonian
  $H(s_j)$ for random time. The number of steps $q$ is obtained so that the final error probability is bounded
  by some given $\varepsilon>0$.}
  \label{fig:RM}
\end{figure}

The ``randomization method'' (RM) described in Ref.~\cite{boixo:qc2009a}  uses phase randomization to traverse the
eigenpath. The basic idea of the RM is simple: For a Hamiltonian path $\{H(s) \}$,
we choose a discretization $0<s_1<s_2<\ldots<s_q=1$ that depends on
the final-state preparation error. At the $j$ th step of the RM,  
we evolve with the constant Hamiltonian $H(s_j)$ for random time $t_j$, which is
drawn according to a specific distribution that depends on $\Delta(s_j)$, the gap at that step, and the error.
A common example is to sample $t_j$ from a normal distribution of zero mean and width (standard deviation)
of order $1/\Delta(s_j)$.
Evolution randomization  will induce phase cancellation
and a reduction of the {\em coherences} between $\ket{\psi(s_j)}$ and any other
state orthogonal to it (see Secs.~\ref{sec:errors} and~\ref{sec:weakmeasurement}).
In other words, evolution randomization simulates a measurement
of $\ket{\psi(s_j)}$. Then, due to a version of the quantum Zeno effect, a sequence of measurements
of $\ket{\psi(s_1)},\ket{\psi(s_2)},\ldots$ will allow  the preparation of $\ket{\psi(s_q)}$,
with arbitrarily high probability for a proper choice of $s_1,s_2\ldots,s_q$. 
The basic steps of the RM are depicted in Fig.~\ref{fig:RM}; more details are in Secs.~\ref{sec:errors} and~\ref{sec:weakmeasurement}.

The average cost of the RM
is  the number of steps $q$ times the average (absolute) evolution time per randomization step;
the latter is proportional to the inverse spectral gap~\cite{boixo:qc2009a}.
For a uniform parametrization under which $\| \sket{\dot  \psi} \|=L$ for all $s$,  and for error $\varepsilon$,
we obtain $q \propto L^2/\varepsilon$, resulting 
in an optimal average cost of order $L^2/(\varepsilon\Delta)$. An additional logarithmic factor,
coming from Eq.~\eqref{eq:costEPT2}, was needed for the cost analysis of Ref.~\cite{boixo:qc2009a}
if the random times are nonnegative (or nonpositive).

Nevertheless, the given parametrization is not uniform in general.
In this case, the RM is only guaranteed to succeed if
$q=(L^*)^2$, where $L^*$ is an upper bound on $L$ that can be determined from some known properties of $H$. 
As discussed, a standard choice for $L^*$ is the one in Eq.~\eqref{eq:Lbound0},
which results in an overall cost of order $1/\Delta^3$ if we disregard other quantities: the number of points in the discretization is $q \propto \max_s (1/\Delta^2)$.
The goal of this section is to show that the upper bound obtained in Sec.~\ref{sec:pathlength}
can be used to obtain a better discretization for the RM than that of Ref.~\cite{boixo:qc2009a},
resulting in an overall, improved average cost of order $\max_s( 1/\Delta^2)$. We also show how
to avoid the logarithmic correction in the cost by performing a more detailed analysis of 
errors due to randomization, when the random times are nonnegative (or nonpositive).

%%%%%%%%%%%%%%%%%%%%%%%%%%%%%%%%%%%
\subsection{Parametrization errors}
\label{sec:errors}
In this section we analyze the errors due to the discretization, which assumes perfect measurements
of the $\ket{\psi(s)}$ in the RM. Errors from imperfect measurements due to evolution
randomization are analyzed in Sec.~\ref{sec:weakmeasurement}.
We  let $0<s_1<s_2<\ldots <s_q=1$ determine any discretization of the interval
$[0,1]$, where $q$ will be obtained below.
Assuming perfect
measurements of the $\ket{\psi(s_j)}$ and using the union bound, the final error or quantum {\em infidelity} ($1-F$) in the preparation of $\ket{\psi(s_q)}$
can be bounded from above as
\begin{align}
\nonumber
1-F &= 1-\prod_{j=1}^q \cos^2(\alpha_j) \\
\nonumber
& \le \sum_{j=1}^q \sin^2 (\alpha_j)    \; ,
\end{align}
where the `angles' $\alpha_j$ are determined from $\cos \alpha_j = | \! \bra{\psi(s_{j-1})} \psi(s_j) \rangle |$ - see
Fig.~\ref{fig:RM}.
Without loss of generality, we can assume $\sin \alpha_j \ge 0$.

In Appendix~\ref{app:bound1}, we show
\begin{align}
\label{eq:anglebound}
\sin^2 ( \alpha_j) \le (s_j-s_{j-1})  \int_{s_{j-1}}^{s_j} ds \; \|   \sket{\dot \psi}\|^2  \; ,
\end{align}
for a differentiable path. If we choose a discretization so that $s_j=j \, \delta  \mspace{-1mu} s$, 
the choice 
$\delta  \mspace{-1mu}  s\le   \varepsilon/ \int_0^1 ds \| \partial_s \ket{\psi(s)} \|^2$
suffices to guarantee a final infidelity bounded by $\varepsilon$;
that is
\begin{align}
\label{eq:fidelity1}
\sum_{j=1}^q \sin^2 (\alpha_j) \le \varepsilon \; .
\end{align}
We can then use the main result of Sec.~\ref{sec:pathlength}  and Eq.~\eqref{eq:anglebound} to show 
\begin{align}
\nonumber
\delta  \mspace{-1mu} s =   \frac { \varepsilon} {(L^*)^2} \; .
\end{align}
This bound assumes that $\ket{\psi}$ is the ground state of $H$.
The number of points in the discretization is then
\begin{align}
\label{eq:pointsRM}
q = \frac 1 {\delta  \mspace{-1mu}  s}= \frac{ \int_0^1 ds   \bra \psi \ddot H - \ddot E \ket \psi/(2 \Delta)} {\varepsilon} \; ,
\end{align}
which is of order $\max_s (1/\Delta)$ if we ignore other quantities.
It follows that the overall, average cost of the RM is of order $\max_s (1/\Delta^2)$,
implying a better gap dependence than the one obtained in Ref.~\cite{boixo:qc2009a}.
In the following section we show how the measurements can be simulated and approximated
by evolution randomization.

%%%%%%%%%%%%%%%%%%%%%%%%%%%%%%%%%
\subsection{Imperfect measurements}
\label{sec:weakmeasurement}
A perfect, projective measurement of $\ket{\psi}$
is one that transforms all coherences  between $\ket \psi$ and
its orthogonal complement to 0. That is, if $\rho$ denotes
the density matrix after the perfect measurement, then $\bra \psi \rho \sket{\psi^\perp}=0$
for all states $\sket{\psi^\perp}$ satisfying $\bra \psi \psi^\perp \rangle=0$.
In the RM of Ref.~\cite{boixo:qc2009a},
we showed that a perfect measurement can only be simulated if the random evolution time $t$
is drawn according to a distribution in which $t \in (-\infty,\infty)$.
If $t$ can only be nonnegative (or nonpositive), the coherences are only reduced by a multiplicative factor $\varepsilon'>0$;
 that is, the simulated measurement is imperfect or {\em weak}. To achieve overall error of order $\varepsilon$ in the preparation
 of the final eigenstate
 due to imperfect measurements,  in Ref.~\cite{boixo:qc2009a} we chose $\varepsilon'=\varepsilon/q$,
 which easily follows from  an union-like bound for a sequence of quantum operations.
This introduces an additional cost to the RM
 given by a multiplicative factor of order $\log (q/\varepsilon)$ [Eq.~\eqref{eq:costEPT1}], which can be large
if $q \gg 1$. Nevertheless,   we now present an improved error analysis of the RM than that
of Ref.~\cite{boixo:qc2009a}, and show that if the imperfect measurements are such that $\varepsilon'$ is a constant
independent of $\varepsilon$, 
an overall error of order $\varepsilon$ can still be achieved. This results in an improved cost for the RM: the $\log (q/\varepsilon)$ overhead is   unnecessary.

 To demonstrate the improved scaling,   it is convenient to define $\rho_j$ as the state, or density matrix, at the $j$ th step of the RM ($j=0,1,\ldots,q$); that is, the state after the randomized evolution with $H(s_j)$. 
 Without loss of generality, we write
 \begin{align}
 \nonumber
 \rho_j = \Pr(j) \ket{\psi(s_j)} \! \bra{\psi(s_j)} + (1-\Pr(j))\rho_j^\perp + \\
 \nonumber
  + \ket{\xi_j} \! \bra{\psi(s_j)} +\ket{\psi(s_j)} \! \bra{\xi_j} \; ,
 \end{align}
where $\Pr(j) = \bra{\psi(s_j)} \rho_j \ket{\psi(s_j)} $ is the probability of $\ket{\psi(s_j)}$ in $\rho_j$ (i.e., the fidelity).
 $\rho_j^\perp$ is a density matrix with support orthogonal to $\ket{\psi(s_j)}$ so that $\rho_j^\perp \ket{\psi(s_j)}=0$.
The (unnormalized) state $\sket{\xi_j}$ is also orthogonal to $\ket{\psi(s_j)}$  and denotes the {\em coherences}
between $\ket{\psi(s_j)}$ and its orthogonal complement. The  norm of $\sket{\xi_j}$ denotes a coherence factor:
\begin{align}
\nonumber
c_j = \| \ket{\xi_j} \| \; .
\end{align}
The main goal of the RM is to simulate measurements by  keeping $c_j$ sufficiently small via phase or evolution randomization.

At the $j+1$ th step, we evolve with $H(s_{j+1})$ for a random time drawn from some distribution $f(t)$.
Then,
\begin{align}
\rho_{j+1}=\int dt \; e^{-i H(s_{j+1})t} \rho_j e^{i H(s_{j+1})t} \; .
\end{align}
Since evolving with $H(s_{j+1})$ leaves the eigenstate
 $\ket{\psi(s_{j+1})}$ invariant (up to a global phase), we have 
 \begin{align}
 \nonumber
 \Pr(j+1) &= \bra{\psi(s_{j+1})} \rho_{j+1} \ket{\psi(s_{j+1})} \\
 \nonumber &=\bra{\psi(s_{j+1})} \rho_j \ket{\psi(s_{j+1})} \; ,
 \end{align}
 with $\ket{\psi(s_{j+1})}=\cos \alpha_{j+1} \ket{\psi(s_{j})} + \sin \alpha_{j+1} \ket{\psi^\perp(s_{j})}$.
Then,
\begin{align}
\label{eq:probstep}
\Pr(j+1) \ge \cos^2 \alpha_{j+1} \Pr(j) - 2  \sin \alpha_{j+1} c_j \; .
\end{align}
Here, we assumed the worst case scenario for which $\sbra{\xi_j} \psi(s_{j+1}) \rangle = -c_j \sin \alpha_{j+1}$ and used $\cos \alpha_{j+1}
\le 1$. In Appendix~\ref{sec:appbound2}, Eq.~\eqref{eq:coherenceboundapp}, we show that if Eq.~\eqref{eq:fidelity1} is satisfied,
\begin{align}
\label{eq:coherencebound}
c_j  \le \frac 1 {1-\varepsilon} (\varepsilon' \sin \alpha_j + \varepsilon'^2 \sin \alpha_{j-1} + \ldots + \varepsilon'^{j} \sin \alpha_{1}) \; .
\end{align}
The factor $\varepsilon'<1$ denotes the reduction in coherence 
due to evolution randomization per step.
That is, 
a random evolution under $H(s_{j+1})$ applied to $\rho_j$ transforms and
reduces the coherences $\sket{\psi(s_{j+1})} \sbra{\psi(s^\perp_{j+1})}$ to
\begin{align}
\nonumber   
&  \int dt \; f(t)  e^{-i H(s_{j+1}) t} \ket{\psi(s_{j+1})} \bra{\psi(s^\perp_{j+1})}e^{i H(s_{j+1}) t}    \; ,
\end{align}
where $\sket{\psi(s_{j+1})^\perp}$ is a normalized state orthogonal to $\sket{\psi(s_{j+1})}$.
Then, we can assume
\begin{align}
\label{eq:epsilon'}
\varepsilon' = \left \|  \int dt \; f(t) e^{i \Delta t} \right \| \;.
\end{align}
where $\Delta \le \Delta(s_{j+1})$.

The RM starts with $\ket{\psi(s_0)}$, so initially $\Pr(0)=1$ and $c_0=0$.
By iteration of Eq.~\eqref{eq:probstep} we obtain
\begin{align}
\label{eq:fidelityRM}
\Pr(q) \ge \prod_{j=1}^q \cos^2( \alpha_j )- 2 \sum_{j=1}^q \sin \alpha_j c_{j-1} \; .
\end{align}
The first term on the {\em rhs} of Eq.~\eqref{eq:fidelityRM} corresponds to the case where all projective measurements 
are  implemented perfectly, i.e., when $c_j=0$ for all $j$. A lower bound to
such term is given by $1-\sum_{j=1}^q \sin^2 (\alpha_j) \ge 1- \varepsilon$, as described in Sec.~\ref{sec:errors}.
Using
Eq.~\eqref{eq:coherencebound}, the second term on the {\em rhs} of Eq.~\eqref{eq:fidelityRM} can be upper bounded by
\begin{align}
\label{eq:geometricdependence}
  \frac 2 {1-\varepsilon} \sum_{j=1}^q \sin \alpha_j (\varepsilon' \sin \alpha_{j-1} + \varepsilon'^2 \sin \alpha_{j-2}+\ldots) \; ,
\end{align}
and using the Cauchy-Schwarz inequality,
\begin{align}
\nonumber
\sum_{j=1}^q \sin \alpha_j \sin \alpha_{j-k}\le  \sum_{j=1}^q \sin^2 (\alpha_j )\le \varepsilon \; .
\end{align}
Then, the fidelity of the RM or probability of success in the preparation of $\ket{\psi(s_q)}$ is
\begin{align}
\label{eq:fidelity2}
\Pr(q) \ge 1 -\varepsilon - \frac {2 \varepsilon \varepsilon'} {(1-\varepsilon)(1-\varepsilon')} \; ,
\end{align}
which follows from summing the geometric series in $\varepsilon'$ in Eq.~\eqref{eq:geometricdependence}.

%%%%%%%%%%%%%%%%%%%%%%%%%%%%%%%%%%%
\subsection{Total cost}
For constant error or infidelity of order $\varepsilon <1$,   it suffices to choose a constant $\varepsilon'$
in Eq.~\eqref{eq:fidelity2}. For example, a common choice for the time distribution is a normal
distribution $f(t)$ with standard deviation of order $1/\Delta$. Since the Fourier transform
of $f(t)$ is  a normal distribution with standard deviation of order $\Delta$, 
Eq.~\eqref{eq:epsilon'} implies a constant upper bound for $\varepsilon'$. 
Then, the average cost per step
of the RM is also of order $ 1/\Delta$. Multiplying this by $q$, the total number
of steps in Eq.~\eqref{eq:pointsRM}, provides an upper bound to the total average cost of the RM given by
\begin{align}
\label{eq:totalcostRM}
\frac{(L^*)^2}{\varepsilon \Delta} \le \kappa' \max_s \frac{\| \ddot H \| + 2 \| \dot H \|} {\varepsilon 2 \Delta^2(s)} \; ,
\end{align}
for general interpolations (Sec.~\ref{sec:generalinterpolations}). $\kappa' \approx \sqrt{2/\pi}$ is also constant~\cite{boixo:qc2009a}. Such an upper bound can be further 
improved for different Hamiltonians
or interpolations as described in Secs. \ref{sec:linearinterpolation} and~\ref{sec:frustrationfree}.
Our result in Eq.~\eqref{eq:totalcostRM} significantly improves upon the result in Ref.~\cite{boixo:qc2009a}, for which 
the average cost in terms of the gap only was of order $\max_s [\log(1/\Delta)/\Delta^3]$.

 %

  %%%%%%%%%%%%%%%%%%%%%%%%%%%%%
  %%%%%%%%%%%%%%%%%%%%%%%%%%%%%
  \section{Applications}
  \label{sec:applications}
Improved bounds on the cost of methods for  eigenpath traversal
may result in  speedups for problems in physics, optimization,
and quantum information. 
In this section, we apply our results to two important examples where polynomial quantum speedups are obtained.

 %%%%%%%%%%%%%%%%%%%%%%%%%%%%% 
\subsection{Preparation of projected entangled pair states (PEPS)}
\label{sec:PEPS}
PEPS, a generalization of matrix product states to
space dimensions higher than one~\cite{RO_MPS_1997,Vidal_MPS_2003}, were conjectured to approximate the ground states of
physical systems with local interactions~\cite{verstraete_peps_2006}. PEPS also arise
in combinatorial optimization  and quantum information problems, and
their preparation  is paramount to solve such problems.
For this reason, methods for the preparation of PEPS on a quantum computer were recently developed~\cite{schwarz_peps_2011,somma_gap_2013}.

An important property of PEPS is that they can be realized as the ground states
of frustration-free Hamiltonians. Then, we can analyze the cost of the RM
for the preparation of PEPS. That is, if $H(s)=\sum_{k=1}^L \Pi_k(s)$ denotes 
a frustration-free Hamiltonian path, using the results of Sec.~\ref{sec:frustrationfree} we obtain a cost for the RM
upper bounded by
\begin{align}
\nonumber
\max_s \frac{\| \ddot H \|} {\varepsilon 2 \Delta^2} \; .
\end{align}
Such a cost can be further improved as follows. A remarkable property of frustration-free Hamiltonians
is that their spectral gap can be amplified by constructing the related Hamiltonian  
\begin{align}
\nonumber
H' =  \sqrt{\| \Pi \|} \sum_{k=1}^L  \sqrt{ \Pi_k} \otimes [ \ket k \bra 0 + \ket 0 \bra k ] \; ,
\end{align}
where $\ket k$, $k=0,1,\ldots L$ are a basis of states of an ancillary system. $H'$ has $\ket \psi \otimes \ket 0$
as eigenstate of eigenvalue 0, and the spectral gap of $H'$ is $\Delta' \ge \sqrt{ \Delta \| \Pi \|}$, where $\| \Pi \| = \max_k 
\| \Pi_k \|$. These properties and the full spectrum of $H'$ was
analyzed in Ref.~\cite{somma_gap_2013}. Then, if we have access to evolutions under the $\sqrt{\Pi_k(s)}$,
the randomized evolution in the RM can be implemented using $H'$ instead, having an average cost
of order $1/\Delta' \propto 1/\sqrt {\Delta \|\Pi \|}$ per step. This implies an overall, average cost for the RM upper bounded by
\begin{align}
\kappa' \max_s \frac{\| \ddot H \|} {\varepsilon 2  \|\Pi\|^{1/2}}\times \frac 1 {\Delta^{3/2}}\; .
\end{align}
Similarly, the cost of other methods for eigenpath traversal~\cite{boixo:qc2010a} for this problem will have an improved cost bounded by
\begin{align}
\label{eq:PEPSbound}
\kappa' \max_s \frac{\sqrt{\| \ddot H \|/2} \times  \log(\sqrt{\| \ddot H/(2 \Delta)}/\varepsilon)} {\varepsilon \|\Pi\|^{1/2} } \times \frac 1 {\Delta} \; .
\end{align}
Equation~\eqref{eq:PEPSbound} follows from Eq.~\eqref{eq:costEPT1} for $c=1$, replacing $\Delta$ by $\Delta'$ and $L$ by $L^*$ as in Eq.~\eqref{eq:pathlengthFF}. The cost is almost linear in $1/\Delta$.

We note that for many frustration-free Hamiltonians, the terms $\Pi_k$ are projectors and $\sqrt{\Pi_k}=\Pi_k$.
Otherwise the $\Pi_k$ may be expressed as a linear combination of projectors, so that the requirement
of having access to evolutions with the $\sqrt{\Pi_k}$ is not strong.

 %%%%%%%%%%%%%%%%%%%%%%%%%%%%%
  \subsection{Quantum Simulated Annealing}
  \label{sec:QSA}
 Simulated annealing is a powerful heuristics for solving combinatorial optimization problems.
 When implemented via Markov-Chain Monte Carlo techniques, it generates a stochastic
 sequence of configurations that converges to the Gibbs distribution determined by the inverse
 {\em temperature} $\beta_q$ and an objective function $E$. For sufficiently large $\beta_q$, 
 the final sequences are sampled from a distribution mostly weighted on those configurations $\sigma$ that minimize $E$.
 The  process is specified by a particular annealing schedule, which consists
 of a finite increasing sequence of inverse temperatures $\beta_0=0<\beta_1<\ldots<\beta_q$.
 The cost of the method is the number of Markov steps required to sample from the desired
 distribution, i.e., $q$. For constant error, such a number can be upper bounded by $\propto \max_\beta 1/\Delta(\beta)$, where $\Delta(\beta)$
 denotes the spectral gap of the stochastic matrix at inverse temperature $\beta$. 
 
 In Ref.~\cite{somma_quantum_2008} we gave a quantum algorithm that allows us to sample from the same 
 distribution as that approached by the simulated annealing method. The 
 quantum algorithm uses the RM to traverse
 a path of states $\ket{\psi(\beta)}$. Here, $\ket{\psi(\beta)}$ is a {\em coherent}
 version of the corresponding Gibbs state, having amplitudes that coincide with
 the square root of the probabilities. That is,
 \begin{align}
 \ket{\psi(\beta)}=\frac 1 {\sqrt{\cal Z}} \sum_\sigma e^{-\beta E[\sigma]/2} \ket \sigma \; ,
 \end{align}
where the sum is over all configurations and ${\cal Z} = \sum_\sigma \exp(-\beta E[\sigma])$ 
is the partition function. 
 
 In more detail, the cost of the quantum method presented in Ref.~\cite{somma_quantum_2008} is of order 
 \begin{align}
 \label{eq:QSAcost}
 \max_{\beta} q \log q/\sqrt{\Delta(\beta)}
 \end{align}
  with
 $q= \beta_q^2 E_M^2 /(4\varepsilon)$ and $E_M$ is the maximum of $|E|$. $\varepsilon$ denotes the overall error probability
 of finding the configuration that minimizes $E$ and $q$ is the number of points
 in the discretization or steps in the RM. As discussed, $q$ is related to the path length
 so that $q \ge L^2/\varepsilon$, with $L=\int_0^{\beta_q} d\beta \| \ket{\partial_\beta \psi(\beta)} \|$
 in this case. In terms of the spectral gap $\Delta(\beta)$, the quantum algorithm
 of Ref.~\cite{somma_quantum_2008} provides a square root improvement over the classical method,
 which is important for those hard instances
where $\Delta(\beta)$ is small.

 We can then use the results in Sec.~\ref{sec:RMbound} to search for a better bound on the path length
 and, ultimately, a reduction on the cost of the RM for this problem. That is,
 instead of using $q$ as above, we replace it by $q^*$, with
 $L^2 \le q^*$ and
 \begin{align}
 \nonumber
 q^* =\frac {\beta_q} \varepsilon \int_0^{\beta_q} d\beta \| \ket{\partial_\beta \psi(\beta)} \|^2 \; ;
 \end{align}
 See Eq.~\eqref{eq:pointsRM}.
  For such $\varepsilon$, $\beta_q$ is of order
 $\log(d/\varepsilon)/\gamma$, where $d$ is the dimension of the configuration space and $\gamma$
   is the difference between the two smallest values
 in the range of $E$ (i.e., the spectral gap of $E$).  
 
 In Appendix~\ref{app:QSA}, Eq.~\eqref{eq:QSAratechange}, we show
\begin{align}
\nonumber
\| \ket{\partial_\beta \psi(\beta)} \|^2 = -\partial_\beta \langle E \rangle /4\; ,
\end{align}
where  $\langle E \rangle$ is the expected (thermodynamic) value of $E$.
Then, we obtain
\begin{align}
\nonumber
q^* = \frac {\beta_q ( \langle E \rangle_{0} - \langle E \rangle_{\beta_q})} {4 \varepsilon} \; .
\end{align}
Without loss of generality, we assume $\langle E \rangle_{0}=0$, as we can always shift the lowest
value of $E$ to satisfy the assumption. In fact, the assumption is readily satisfied for many
problems of interest, such as those where $E$ describes a so-called Ising model. If $\beta_q \gg 1$, then
$\langle E \rangle_{\beta_q} \approx - E_M$ and
\begin{align}
\nonumber
q^* \le \frac {\beta_q   E _M} {4 \varepsilon} \; .
\end{align}
Our improved average cost of the RM for this problem is then 
\begin{align}
 \label{eq:QSAcost2}
T_{\rm QSA}= \kappa' \max_\beta \frac{\beta_q   E _M}{4 \varepsilon \sqrt{\Delta(\beta)}} \; 
\end{align}
($\kappa'$ is a small constant).
Equation~\eqref{eq:QSAcost2}
 has to be contrasted with the worse cost given by Eq.~\eqref{eq:QSAcost}, which in this case is of order
\begin{align}
% \label{eq:QSAcost2}
\nonumber
 \max_\beta \frac{\beta_q^2   E _M^2 \log(\beta_q^2 E_M^2/\varepsilon)}{ \varepsilon \sqrt{\Delta(\beta)}}  \; 
\end{align}
and much larger than $T_{\rm QSA}$ in the large $E_M$ and $\beta_q$ limit.

  %%%%%%%%%%%%%%%%%%%%%%%%%%%%%%%%%%%%%%%%%%%%%%
\section{Conclusions}
\label{sec:conclusions}
We presented a significantly improved upper bound on $L$, the length of the path traversed 
by the continuously-related ground states of a family of Hamiltonians.
Such a bound is approximately the square root of standard and previously
used bounds for $L$ in the literature. It results in an improved average cost of a method for adiabatic state 
transformations based on evolution randomization, which is a simple extension of AQC.
 Specifically, we prove an average cost of order $1/\Delta^2$ for the 
 randomization method, whereas AQC has 
a   proven cost of order $1/\Delta^3$ (i.e., the cost of AQC is upper bounded
by $1/\Delta^3$, disregarding other quantities). Here, $\Delta$ is a bound on the spectral gap of the Hamiltonians.
When the Hamiltonians satisfy 
a certain frustration-free property, the average cost of the randomization method is further
improved to order $1/\Delta^{3/2}$. The gap $\Delta$ is very small for hard instances
and thus the randomization method is a promising alternative to AQC in these cases,
as it has a proven lower cost.

We also improved the cost of the randomization method
when the simulated measurements are imperfect. We showed
that if evolution randomization induces a weak measurement, where the coherences
are reduced by a constant, multiplicative factor (e.g., by 1/3), then the eigenstate
of the final Hamiltonian is still prepared at small, bounded error probability.
Previous analysis for the randomization method required a reduction on the coherences
that depended on the path length.

The randomization method outperforms AQC in certain instances (e.g., Rabi oscillations).
Nevertheless, it remains open to show how generic the advantages of the randomization method
over AQC are. To understand this problem better, for example, one needs to devise other instances
where AQC has a cost dominated by $1/\Delta^3$, so that the cost of AQC is strictly higher
than that of the randomization method.   Perhaps our most important contribution is a method
for eigenpath traversal that has a {\em proven} lower cost than that provided by quantum
adiabatic approximations~\cite{jansen_bounds_2007,jordan_thesis_2008,regev_quantum_2004,lidar_adiabatic_2009}, since
rigorously improving the latter cost in terms of the gap, even for simple cases (e.g., linear interpolations),
does not seem feasible.

Finally, the improved bound on $L$ can also be used to improve the cost of other methods
for eigenpath traversal such as that in Ref. \cite{boixo:qc2010a}. For the most efficient and 
known method for eigenpath traversal in the literature, our bound on $L$ implies a cost of order $1/\Delta^{3/2}$
for general Hamiltonians and order $1/\Delta$ for Hamiltonians that satisfy the frustration free property.

 %%%%%%%%%%%%%%%%%%%%%%%%%%%%%%%%%%%%%%%%%%%%%%
 % \section{Conclusion}

 %%%%%%%%%%%%%%%%%%%%%%%%%%%%%%%%
 \section{ Acknowledgements}
H.-T.C. acknowledges  support from the National Science Foundation through the CCF program. GX and RS
acknowledge support from  AFOSR through grant number FA9550-12-1-0057. RS thanks
Sandia National Laboratories, where the initial ideas of this work were developed.
 Sandia National Laboratories is a multi-program laboratory managed and operated by Sandia Corporation, a wholly owned subsidiary of Lockheed Martin Corporation, for the U.S. Department of Energy's National Nuclear Security Administration under contract DE-AC04-94AL85000.
We thank Sergio Boixo, Andrew Daley and Andrew Landahl for  discussions.

%%%%%%%%%%%%%%%%%%%%%%%%%%%%%%%%%%%%%%%%%%%%%%%%
\begin{appendix}

%%%%%%%%%%%%%%%%%%%%%%%
\section{A bound on $\| \sket{\dot \psi}\|$}
\label{app:rateofchange}
From $H \ket \psi=0$, where $\ket \psi$ is the ground state, we obtain
\begin{align}
\nonumber
\sket{\dot \psi} =  -H^{-1} \dot H \ket \psi \; .
\end{align}
$H^{-1}$ denotes the operator that is inverse to $H$ in the
subspace orthogonal to $\ket{\psi}$. We assume the existence of $\dot H$
with $\| \dot H \| <\infty$.
Then,
\begin{align}
\nonumber
\| \sket{\dot \psi} \| ^2 &= \bra \psi \dot H H^{-2} \dot H \ket \psi \\
\nonumber
& \le \frac 1 \Delta \bra \psi \dot H H^{-1} \dot H \ket \psi \\
\label{eq:dotpsi}
& =  \frac {-1} \Delta \bra \psi \dot H  \sket{\dot \psi}
\end{align}
where we used Cauchy-Schwarz
and the assumption that $H \ge 0$.
In addition,
\begin{align}
%\label{eq:Lbound3}
\nonumber
\dot H \sket{\dot \psi} = - \frac 1 2 [\ddot H \ket \psi + H \sket{\ddot \psi}] \; ,
\end{align}
and using Eq.~\eqref{eq:dotpsi}
we obtain the desired bound as
\begin{align}
\label{eq:dotpsibound}
\| \sket{\dot \psi} \|^2 \le \frac {1}{2 \Delta} \bra \psi \ddot H \sket{ \psi} \; .
\end{align}
 This assumes the existence of $\ddot H$ with $\| \ddot H \| <\infty$.

%%%%%%%%%%%%%%%%%%%%%%%
\section{A bound on $\sin \alpha_j$}
\label{app:bound1}
As pointed out, the angles $\alpha_j$ in Sec.~\ref{sec:RMbound} (Fig.~\ref{fig:RM})
can be defined via $\cos \alpha_j =  \bra{\psi(s_{j-1})} \psi(s_j) \rangle \in \mathbb {R}$.
It follows that
\begin{align}
\nonumber
\sin \alpha_j & = \| \ket{\psi(s_{j-1})} - \cos \alpha_j \ket{\psi(s_{j})} \|
\\
\label{eq:app1}
& \le   \| \ket{\psi(s_{j-1})} - e^{i \phi} \ket{\psi(s_{j})} \| \; .
\end{align}
The phase $\phi \in \mathbb{R}$ can be arbitrary.
Next,
we split the interval $[s_{j-1},s_j]$ into $r$ segments of size
$(s_j-s_{j-1})/r$  and define $s_j^n=s_{j-1} +(s_j-s_{j-1}) n/r $, with $n=0,1,\ldots , r$.
The corresponding eigenstates are now $\ket{\psi(s_j^n)}$ and, with no loss of generality, we assume
 $\cos \beta_n = \bra{\psi(s_{j}^{n-1})} \psi(s_j^{n}) \rangle \in \mathbb{R}$. In particular,
 $\ket{\psi(s_j^0)}=\ket{\psi(s_{j-1})}$ and $\ket{\psi(s_j^n)}= e^{i\phi}\ket{\psi(s_{j})}$.

From Eq.~\eqref{eq:app1} we obtain
\begin{align}
\nonumber
\sin \alpha_j & \le \| \sum_{n=0}^{r-1} \left( \ket{\psi(s_{j}^n)} -   \ket{\psi(s_{j}^{n+1})}  \right) \| \\
%\label{eq:app2}
\nonumber
& \le \sum_{n=0}^{r-1} \left \| \ket{\psi(s_{j}^n)} -   \ket{\psi(s_{j}^{n+1})} \right \| \; ,
\end{align}
where we used the triangle inequality. Also,
\begin{align}
\nonumber
\sin \alpha_j & \le \lim_{r \rightarrow \infty} \sum_{n=0}^{r-1}  \frac{\left \| \ket{\psi(s_{j}^n)} -   \ket{\psi(s_{j}^{n+1})} \right \|}{s_j^{n+1}-s_j^n}  
\frac n r (s_j-s_{j-1})  \\
\label{eq:app3}
& \le \int_{s_{j-1}}^{s_j} ds \; \| \sket{\dot \psi } \| \; ,
\end{align}
where the phase of $\ket{\psi}$ must be chosen so that $\langle \dot \psi \ket{\psi} \in \mathbb R$,
and thus $\langle \dot\psi \ket{\psi }=0$ from the normalization condition. The inequality
in Eq.~\eqref{eq:app3} requires existence $\sket{\dot \psi}$, i.e., a differentiable path.
Since
\begin{align}
\nonumber
\int_{s_{j-1}}^{s_j} ds \; \| \ket{\partial_s \psi(s)} \|^2 - \left( \int_{s_{j-1}}^{s_j} ds \; \| \ket{\partial_s \psi(s)} \|  \right)^2 
%&=
%\\
%\nonumber
%= \int_{s_{j-1}}^{s_j} ds \left( \| \ket{\partial_s \psi(s)} \| -  \int_{s_{j-1}}^{s_j} ds' \| \ket{\partial_s' \psi(s')} \| \right)  ^2 & 
\ge 0 \; 
\end{align}
from Cauchy Schwarz, we obtain the desired bound as
\begin{align}
\nonumber
\sin \alpha_j \le \left( \int_{s_{j-1}} ^{s_j } ds \; \| \sket{\dot \psi } \|^2 \right)^{1/2}\; .
\end{align}
 
%%%%%%%%%%%%%%%%%%%%%%%%%%%%%%%%%
 \section{A bound on the coherences}
 \label{sec:appbound2}
 As explained in Sec~\ref{sec:weakmeasurement}, we let $\rho_j$ be the density matrix
 for the state after the randomized evolution with $H(s_j)$, i.e.,
 the state output at the $j$ th step of the randomization method:
  \begin{align}
 \nonumber
 \rho_j = \Pr(j) \ket{\psi(s_j)} \! \bra{\psi(s_j)} + (1-\Pr(j))\rho_j^\perp + \\
 \nonumber
  + \ket{\xi_j} \! \bra{\psi(s_j)} +\ket{\psi(s_j)} \! \bra{\xi_j} \; ,
 \end{align}
 The coherence factor is defined as
 \begin{align}
 \nonumber
 c_j = \| \ket{ \xi_j} \| =\| P_j^\perp \rho_j \ket{\psi(s_j)} \| \; ,
 \end{align}
  where $P_j^\perp = \one - \ket{\psi(s_j)} \bra{\psi(s_j)}$ is the a projector onto the subspace orthogonal to $\ket{\psi(s_j)}$.
  The coherence factor at the $j+1$ th step is then
   \begin{align}
 %  \label{eq:app4}
 \nonumber
 c_{j+1} & = \| \ket{\xi_{j+1}} \| \\ \nonumber
 &=\| P_{j+1}^\perp \rho_{j+1} \ket{\psi(s_{j+1})} \| \\
 \nonumber
 & = \| P_{j+1}^\perp \int dt \; f(t) e^{-iH(s_{j+1})t} \rho_{j} e^{iH(s_{j+1})t} \ket{\psi(s_{j+1})} \| \; ,
%  \le \varepsilon' \| P_{j+1}^\perp \rho_j \ket{\psi(s_{j+1})} \| \; ,
 \end{align}
 where $f(t)$ is the distribution for the random time at that step.
 Since $e^{iH(s_{j+1})t}$ leaves $\ket{\psi(s_{j+1})}$ invariant (up to a global phase)
 and
 \begin{align}
 \nonumber
\left \| \int dt \; f(t) e^{-iH(s_{j+1})t} \sket{\bar \psi^\perp(s_{j+1})} \right \| \le \varepsilon'
 \end{align}
 for any unit state $\sket{\bar \psi^\perp(s_{j+1})}$ orthogonal to $\ket{ \psi(s_{j+1})}$,
 we arrive at
 \begin{align}
   \label{eq:app4}
 c_{j+1} \le \varepsilon' \| P_{j+1}^\perp \rho_j \ket{\psi(s_{j+1})} \| \; .
 \end{align}
 The factor $\varepsilon'<1$ was defined in Eq.~\eqref{eq:epsilon'}, and is the Fourier transform
 of $f(t)$ at $\Delta \le \Delta(s_{j+1})$.
% 
% 
% since the effect of evolution randomization with $H(s_{j+1})$ is to reduce the entries of $\rho_j$
% associated with $\ket{\psi(s_{j+1})}$ and its orthogonal subspace.
 
 We now bound the {\em rhs} of Eq.~\eqref{eq:app4}.
 Without loss of generality, we write $\ket{\psi(s_{j+1})} = \cos \alpha_{j+1} \ket{\psi(s_{j})} + \sin \alpha_{j+1} \ket{\psi^\perp(s_{j})}$,
 and obtain
   \begin{align}
   \nonumber 
 c_{j+1} \le \varepsilon' \left[ \cos \alpha_{j+1} \| P_{j+1}^\perp  \left( \Pr(j)  \ket{\psi(s_{j})} + \ket{\xi_j} \right) \|  \right. + \\
 %\label{eq:app5} 
 \nonumber 
 \left.
 + \sin \alpha_{j+1} \|  P_{j+1}^\perp \rho_j \ket{\psi^\perp(s_{j})} \| \right] \; ,
 \end{align}
  where we used the triangle inequality and $\rho_j \ket{\psi(s_j)} = \Pr(j) \ket{\psi(s_j)} + \ket{\xi_j}$.
  By definition, $\sin \alpha_{j+1} = \| P_{j+1}^\perp  \ket{\psi(s_{j})}\| $. Also,
  \begin{align}
  \nonumber
  \rho_j & \ket{\psi^\perp(s_{j})} = \\
  \nonumber
  & = (1-\Pr(j)) \rho_j^\perp \ket{\psi^\perp(s_{j})} + \ket{\psi(s_j)} \! \bra{\xi_j} \psi^\perp(s_{j})\rangle \; .
  \end{align}
By using Cauchy-Schwarz and the triangle inequalities, we obtain
\begin{align}
\nonumber
c_{j+1} \le \varepsilon' \left[ \cos \alpha_{j+1} \Pr(j) \sin \alpha_{j+1} + \cos \alpha_{j+1} c_j + \right. \\
\nonumber 
\left. + \sin \alpha_{j+1} (1-\Pr(j)) + \sin^2( \alpha_{j+1}) c_j  \right] \; ,
\end{align}
and thus
\begin{align}
\label{eq:iterationstep}
c_{j+1} \le \varepsilon' \left[ \sin \alpha_{j+1} + (1+\sin^2 (\alpha_{j+1})) c_j \right] \;.
\end{align}
Because the initial state (step 0) is exactly $\ket{\psi(s_0)}$, we have $c_0=0$ and,
by iteration of Eq.~\eqref{eq:iterationstep},
\begin{align}
\nonumber
c_{j+1} \le \varepsilon' \sin \alpha_{j+1} + (\varepsilon')^2 (1+ \sin^2 (\alpha_{j+1})) \sin \alpha_{j} + \ldots \\
\nonumber
\ldots + (\varepsilon')^{j+1} (1+ \sin^2( \alpha_{j+1})) \ldots (1+ \sin^2 (\alpha_2))\sin \alpha_{1} \; .
\end{align}
In order to relate $\varepsilon'$ with the error coming from the discretization (perfect measurements),
we recall the condition
\begin{align}
\nonumber
\sum_{j=1}^q \sin^2 (\alpha_j )\le \varepsilon \; 
\end{align}
of Eq.~\eqref{eq:fidelity1}.
Then,
\begin{align}
\nonumber
&\prod_{j=i}^q (1+\sin^2( \alpha_j)) \le \prod_{j=1}^q (1+\sin^2 (\alpha_j)) \\
\nonumber
& \le 1 + \sum_{j=1}^q \sin^2( \alpha_j)+ \left( \sum_{j=1}^q \sin^2 (\alpha_j) \right)^2 + \ldots \\
\nonumber
& \le \sum_{j \ge 0} \varepsilon^j= 1/(1-\varepsilon)\; ,
\end{align}
where the last inequality is due to the geometric series.
Then,
\begin{align}
\label{eq:coherenceboundapp}
c_j \le \frac 1 {1-\varepsilon} (\varepsilon' \sin \alpha_j + \varepsilon'^2 \sin \alpha_{j-1} + \ldots + \varepsilon '^{j} \sin \alpha_{1}) \; ,
\end{align}
which is the desired bound.

%%%%%%%%%%%%%%%%%%%%%%%%%%%%%%%%%%%%%
\section{Eigenstate change in QSA}
\label{app:QSA}
By definition, the eigenstate path in QSA is determined by
 \begin{align}
 \nonumber
 \ket{\psi(\beta)}=\frac 1 {\sqrt{\cal Z}} \sum_\sigma e^{-\beta E[\sigma]/2} \ket \sigma
 \end{align}
where $0 \le \beta \le \beta_q$, $E[\sigma] \in \mathbb R$
is the value of the objective function for (classical) configuration $\sigma$, and ${\cal Z}=\sum_{\sigma} e^{-\beta E[\sigma]}$ is the partition function.
Then, it is simple to show
 \begin{align}
  \label{eq:QSAratechange0}
\ket{ \partial_\beta \psi(\beta)}=\frac 1 2 \left [\langle E \rangle \ket{\psi(\beta)}- \frac 1 {\sqrt{\cal Z}} \sum_\sigma E[\sigma]   e^{-\beta E[\sigma]/2} \ket \sigma \right] \; ,
 \end{align}
where
\begin{align}
\nonumber
\langle E \rangle = \frac 1 {\cal Z} \sum_\sigma E[\sigma]   e^{-\beta E[\sigma]}
\end{align}
is the expected (thermodynamic) value of $E$ at inverse temperature $\beta$. Because $\{\ket \sigma \}$ is an orthogonal
basis, Eq.~\eqref{eq:QSAratechange0} gives
 \begin{align}
 \nonumber
\| \ket{ \partial_\beta \psi(\beta)} \|^2 &= \frac 1 4 \sum_\sigma (\langle E \rangle -E[\sigma])^2 \times \frac{e^{-\beta E[\sigma]}}{\cal Z} \\
\nonumber
& = \frac 1 4 \left( \langle E^2 \rangle - \langle E \rangle^2 \right) \; ,
 \end{align}
 relating the rate of change of the state with the thermodynamic fluctuations of $E$.
In addition,
\begin{align}
\nonumber
\partial _ \beta \langle E \rangle   &=\partial _ \beta  \frac 1 {\cal Z} \sum_\sigma  E[\sigma] e^{-\beta E[\sigma]}  \\
\nonumber
& = \frac{-\partial_\beta \cal Z}{{\cal Z}^2}\sum_\sigma  E[\sigma] e^{-\beta E[\sigma]} - \frac{1}{\cal Z}\sum_\sigma  E^2[\sigma] e^{-\beta E[\sigma]} \\
\nonumber
& = \langle E \rangle^2 - \langle E^2 \rangle \;,
\end{align}
and then
\begin{align}
 \label{eq:QSAratechange}
\| \ket{ \partial_\beta \psi(\beta)} \|^2 &= -\frac{\partial_\beta \langle E \rangle} 4 \; .
\end{align}

\end{appendix}

% \bibliography{EPT-Bounds}

\begin{thebibliography}{45}
\bibitem [{Not()}]{NoteEPT-bounds}%
  \BibitemOpen
  \href@noop {} {}\bibinfo {howpublished} {With no loss of generality, we can
  assume that the eigenvalue is zero so that $H \ket{\psi }=0$. Then,
  $\sket{\dot \psi}=(-1/H)\dot H \ket \psi$, where $1/H$ has support on the
  space orthogonal to $ \ket{\psi}$ only.}\BibitemShut {Stop}%
\end{thebibliography}
%

\end{document}